\documentclass[letterpaper]{aa}  
\usepackage{natbib}
\bibpunct{(}{)}{;}{a}{}{,}
\usepackage[figuresright]{rotating}
\usepackage{graphicx}
\usepackage{txfonts}
\begin{document}
   \title{The CFHTLS Strong Lensing Legacy Survey:}
   \titlerunning{CFHTLS-SL2S overview}
   \authorrunning{Cabanac et al.}
   \subtitle{I. Survey overview and T0002 release sample}
   \author{R.~A.~Cabanac\inst{1}, C.~Alard\inst{2},
   M.~Dantel-Fort\inst{2}, B.~Fort\inst{3}, R.~Gavazzi\inst{4},
   P.~Gomez\inst{5}, J.P.~Kneib\inst{6}, O.~Le~F\`evre\inst{6},
   Y.~Mellier\inst{3,2}, R.~Pello\inst{4}, G.~Soucail\inst{4},
   J.F.~Sygnet\inst{3}, D.~Valls-Gabaud\inst{1,4,7} 
\thanks{Based on observations obtained with MegaPrime/MegaCam, a joint
project of CFHT and CEA/DAPNIA, at the Canada-France-Hawaii Telescope
(CFHT) which is operated by the National Research Council (NRC) of Canada,
the Institut National des Sciences de l'Univers of the Centre National
de la Recherche Scientifique (CNRS) of France, and the University of
Hawaii. This work is based in part on data products produced at TERAPIX
and the Canadian Astronomy Data Centre as part of the Canada-France-Hawaii
Telescope Legacy Survey, a collaborative project of NRC and CNRS.}
          }

   \offprints{R. Cabanac, cabanac@cfht.hawaii.edu}

   \institute{
      Canada-France-Hawaii Telescope, 
      65-1238 Mamalahoa Hw., Kamuela, Hawaii 96743, USA
         \and
      Observatoire de Paris, LERMA, CNRS-UMR8112, 
      61 avenue de l'Observatoire, 75014 Paris, France
         \and
      Institut d'Astrophysique de Paris, CNRS-UMR7095 and Universit\'e Pierre et Marie Curie, 
      98 bis boulevard Arago, 75014 Paris, France
         \and
      Laboratoire d'Astrophysique de Toulouse-Tarbes, CNRS-UMR5572 and Universit\'e Paul 
      Sabatier Toulouse III, \\
      14 avenue Edouard Belin, 31400 Toulouse, France
         \and
      Gemini Observatory Northern Operations Center, 
      670 N. A'ohoku Place, Hilo, Hawaii 96720, USA
         \and
      Laboratoire d'Astrophysique de Marseille, 
      traverse du Siphon BP8, 13376 Marseille Cedex 12, France
         \and
      Observatoire de Paris, GEPI, CNRS-UMR8111, 
      5 place Jules Janssen, 92195 Meudon Cedex, France
}

   \date{}


 
  \abstract
{}
{We present data from the CFHTLS Strong Lensing Legacy Survey 
(SL2S). Due to the unsurpassed combined depth, area and image 
quality of the Canada-France-Hawaii Legacy Survey it is becoming 
possible to uncover a large, statistically well-defined sample of 
strong gravitational lenses which spans the dark halo mass spectrum 
predicted by the concordance model from  galaxy to cluster haloes.}
{We describe the development of several automated procedures to find
strong lenses of various mass regimes in CFHTLS images.}
{The preliminary sample of about 40 strong lensing candidates 
discovered in the CFHTLS T0002 release, covering an effective field of view 
of 28 deg$^2$ is presented. These strong lensing systems were discovered
using an automated search and consist mainly 
of gravitational arc systems with splitting angles 
between 2 and 15 arcsec. This sample shows for the first time 
that it is possible to uncover a large population 
of strong lenses from galaxy groups with typical halo masses of
about $10^{13}h^{-1}M_\odot$. We discuss the future 
evolution of the SL2S project and its main scientific aims for the 
next 3 years, in particular our observational 
strategy to extract the hundreds of gravitational rings also 
present in these fields.}
   {}

   \keywords{gravitational lensing -- surveys -- galaxy evolution -- 
galaxy mass -- dark matter}

   \maketitle
%

\section{Introduction}

The observation of gravitational lensing effects produced
by mass concentrations is a powerful tool to directly probe
dark matter haloes and their interplay with visible mass 
\citep{schneider92,blandford92,miraldaescude92,mellier02,kneib03}.
Both strong and weak lensing regimes are therefore widely used to explore the 
dark matter distribution properties, either from observations of individual 
cases \citep{warren96,cabanac05,willis06}
or from statistical analyses of large samples of lensed galaxies 
or quasars 
\citep[for a comprehensive review][]{mellier05}. 

On galaxy scales, ongoing lens surveys are now providing reliable 
descriptions of galaxies and a clearer understanding of the key 
issues regarding the star and dark matter distributions.  
As pointed out by \citet{rusin03} and \citet{rusin05}, using a large 
sample of well-studied strong lenses, it is possible to 
describe galaxy structure and the transition between the inner 
stellar matter-dominated and the outer dark matter-dominated galaxy haloes, 
without being sensitive to the mass-sheet degeneracy. 
It complements well the galaxy-galaxy lensing methods that explore galaxy 
haloes on much larger scales \citep{hoekstra04,seljak05}.
Detailed lens studies have led to an observational technique 
based on spectroscopic selection of compact lensing galaxy candidates 
\citep{hall00,hewett00} that was fully exploited 
by \citet{bolton06,treu05,koopmans06} and 
\citet{willis06}. Using the large spectroscopic data
base of the SDSS, \citet{bolton06} have identified and
studied a first set of 20 rings (which are merged compact multiple arc 
systems) among a total sample of 120 candidates. 
The first analysis of this Sloan Lens ACS (SLACS) data 
has provided a better description of the structural parameters 
of isolated galaxy lenses at low redshift
than multiply-imaged quasars \citep{treu05,koopmans06}. 
Moreover, their study of the fundamental plane of E/SO galaxies 
improves on ealier work \citep{kochanek00} by combining
a standard dynamical analysis with the strong lens constraints. This
allows them to break the degeneracy between anisotropy of the stellar
velocity tensor and the lens gravitational potential \citep[for a similar
approach][]{miraldaescude95}.  They also showed that the lens galaxies of
the SLACS sample nicely follow the E/S0 fundamental plane, being only
slightly skewed towards the more massive objects. However, the SLACS is
limited to small rings ($<$3\arcsec) by the aperture of the spectrograph 
fibers and to nearby lenses  ($z_{max}(lens)<0.5$, 
$z_{max}(source)<0.8$; cutoff of the SDSS 
spectroscopic follow-up, and SLACS selection criteria). 
An extension of the method to larger redshifts should enlarge the 
sample significantly and will benefit from the higher efficiency of 
strong lensing at redshift $\sim 0.5$ for galaxy sources at redshift 
above 1. Such ongoing efforts by \citet{willis06} push
$z_{max}(source)<1.3$.

Because clusters of galaxies are more complex systems than galaxies alone, 
it is not yet clear to what extend systematics, projection and selection 
effects hamper a reliable description of cluster size haloes. 
While the number of cluster-size haloes can in principle be derived from weak 
lensing studies \citep{hetterscheidt05}, the detailled description 
of the cluster halo structures and of their light versus mass distribution 
properties is still uncertain. Ground-based+HST observations and  
strong+weak lensing analyses of individual or samples of 
clusters of galaxies seem to indicate that more complex radial profiles 
(NFW like or power law with a flat core) than singular isothermal 
are required to fit the lensing data \citep{kneib03,gavazzi03,broadhurst05b}. 
However for giant arcs in clusters of galaxies, large optically selected samples 
of strong lensing groups of galaxies are not available yet.
The Red-sequence Cluster Survey (Gladders et al 2003) is a first
attempt to systematically find strong lensing around clusters and groups of
galaxies. But their relatively low detection sensitivity has lead
to the discovery of only eight cluster-like structures at $z>0.64$ over
a 90 sq.degree field.

A key issue is to understand the transition between galaxy-scale 
to cluster-scale halo structures. Quasar lenses and gravitational arcs 
have mostly probed two regimes of halo masses: galaxies and clusters of 
galaxies, but bring only weak constraints on the intermediate mass range 
$(10^{12}-10^{14}$ M$_\odot$) which is important for the assembly 
of large scale structures \citep{kochanek01b,grant04,fassnacht05, matthews05,oguri06}. The study of groups of galaxies in the CNOC survey 
using weak lensing by \citet{parker05} yielded the first constraints of 
their averaged mass-to-light ratios but nothing on their inner structures or 
on whether groups are self-similar. 
%

In summary, no homogeneous sample of strong lenses have been built 
so far that covers the full dark matter halo mass spectrum 
because of the lack of a large, deep sky survey with a sub-arc-second seeing.
We will demonstrate here that with the CFHT Legacy Survey 
(CFHTLS), due to its combined depth, 
area and image quality of the data, we are able to find 
strong lensing systems around a wide mass spectrum of structures.
Indeed, the three $7\times7$-deg$^2$ wide patches together with the four 
$1\times1$-deg$^2$ deep patches of CFHTLS allow us to build up 
a large sample of strong galaxy-, group-, and cluster-scale lenses 
with a well-defined selection function and sampling variance, 
as well as to explore halo properties at different depths and redshifts.

In order to do this, a set automated procedures has been developed 
to detect various types of strong lensing events. It has been successfully 
tested on the T002 release. We present in this paper a preliminary sample built
with these selection procedures. Although these procedures are
not yet fully optimized, we can nevertheless uncover within the CFHTLS
a large population of ``group lenses'', a new class of 
lenses with multiple image separation of 2 to 7\arcsec. 
We also show that it is possible to implement a 
dedicated procedure to recover gravitational rings with Einstein radii 
below 2\arcsec.  The CFHTLS Strong Lensing Legacy 
Survey (SL2S) should allow us to extract the whole lensing mass 
spectrum from galaxies to clusters of galaxies, in a 
homogeneous and statistically well-defined procedure.

The paper is organized as follow: Section 2 summarizes the present 
state of the CFHTLS and the data used in this preliminary study. 
Section 3 presents the SL2S project itself with a brief 
discussion of the automated search procedures, as well as the first 
results obtained for group lenses. A full description of the
selection procedures will be provided elsewhere. Sections 4 and 5
discuss the future of the project.

Throughout the paper we use a flat $\Lambda$CDM cosmology
($\Omega_m=0.3, \ \Omega_\Lambda=0.7$), all observables are
computed with $H_0 = 70 \ h$ km s$^{-1}$ Mpc$^{-1}$ and
magnitudes are given in the $AB$ system unless specified
otherwise.


\section{The Canada-France-Hawaii Telescope Legacy Survey \label{CFHTLS}}
\subsection{Description}
The Canada-France-Hawaii Telescope Legacy Survey (CFTHLS) is a
major photometric survey of more than 450 nights over 5 years
(started on June 1st, 2003) using the wide field imager MegaPrime
which covers $\sim$1 square degree on the sky, with a pixel size of
0.186\arcsec . The project is comprehensively described in {\tt
http://www.cfht.hawaii.edu/Science/CFHLS/} and links therein. The CFHTLS
has two components aimed at extragalactic studies: a very Deep component
made of 4 pencil-beam fields of 1\,deg$^2$ and a Wide component made of
3 mosaics covering 170\,deg$^2$ in total, both in 5 broadband filters. The
data are pre-reduced at CFHT with the Elixir pipeline\footnote{\tt
http://www.cfht.hawaii.edu/Instruments/Elixir/} which removes the
instrumental artefacts in individual exposures. The CFHTLS images are
then evaluated, astrometrically calibrated, photometrically 
inter-calibrated, resampled and stacked by the Terapix group at the 
Institut d'Astrophysique de Paris (IAP) and finally archived at the Canadian 
Astronomy Data Centre (CADC). Terapix also provides weightmap images,
quality assessements meta-data for each stack as well as mask files 
that mask straylight, saturated stars and defects on each image.
The preliminary SL2S sample presented here is based on the T0002 release
(July 2005), corresponding to data obtained between June 1st, 2003
and Nov. 22nd, 2004.  A detailed description of this release
is given at the Terapix web site {\tt http://terapix.iap.fr} and
the Terapix T0002 release document (Mellier et al 2005: 
{\tt http://terapix.iap.fr/IMG/pdf/Newterapixdoc.pdf}). 
The T0002 release includes 40 stacked fields in the Wide survey 
observed in broadband $g'$, $r'$ and $i'$ filters, and a stack of the 
4 Deep fields in the 5 bands, for a total area of 44 deg$^2$, or 
ca. 28 deg$^2$ of unmasked area, i.e. area not contaminated by instrumental 
artefacts from bright stars (internal reflections, bleeding).  
Table \ref{cfhtls_t2} summarizes the main characteristics of the data used in 
the paper. The 4 Deep fields are much deeper, with an average seeing 
ranging from 1.0\arcsec\ in $u^*$ to 0.85\arcsec\ in $z'$. 
The Wide survey is presently available in 3 filters only with an average
seeing of 1.0\arcsec\ in $g'$ and 0.9\arcsec\ in $r'$ and $i'$.
But not surprisingly, the Wide survey is more suited to our 
strong lensing selection processes, because of its wide angular
coverage.

\begin{table}
\caption{CFHTLS: Terapix T0002 release (July 2005)\label{cfhtls_t2}}
\begin{tabular}{lccccc}
\hline
{\bf Deep fields} & \multicolumn{5}{c}{\bf Magnitudes Limits$^1$}\\
& $u^*$&$g'$&$r'$&$i'$&$z'$\\
\noalign{\smallskip}
D1 & 26.4 & 26.3 & 26.1 & 25.9 & 24.9\\
D2 & 26.1 & 25.8 & 25.8 & 25.4 & 24.3\\
D3 & 25.8 & 26.3 & 26.3 & 25.9 & 24.7\\
D4 & 26.2 & 26.3 & 26.3 & 25.7 & 24.9\\
\noalign{\medskip}
{\bf Wide fields} & \multicolumn{5}{c}{\bf Average Magnitudes Limits$^1$}\\
\noalign{\medskip}
W1  & - & 25.5 & 24.5 & 24.5 & -\\
W2  & - & 25.4 & 24.7 & 24.6 & -\\
W3  & - & 25.7 & 25.0 & 24.6 & -\\
\noalign{\smallskip}
W1 area (unmasked) & \multicolumn{5}{c}{20 (11.4) deg$^2$}\\
W2 area (unmasked) & \multicolumn{5}{c}{~8 ~~(5.0) deg$^2$}\\
W3 area (unmasked) & \multicolumn{5}{c}{12 ~(9.0) deg$^2$}\\
\noalign{\smallskip}
\hline
\end{tabular}

$^1$ 50\% completeness limit in $AB$ mag ($AB$ to Vega $u^*-0.35$, $g'+0.09$, $r'-0.17$, 
 $i'-0.40$, $z'-0.55$)
\end{table}

\subsection{Strong lensing number predictions}
In this paper we call a \emph{ring} any compact system of arcs
where multiple images merge into a single ring-like 
image surrounding the deflector; all other types of multiple systems
are either called (giant) \emph{arcs} or \emph{arclets}.
The previous observations of QSO lenses and the results from the
most recent simulations (lensing optical depths) give good estimates of
number densities of lenses associated at various deflecting halo masses
(i.e. for splitting angles) \citep[ Fig.~9 and 10, and Table~1]{oguri06}. 
Typically, for any giant arc detected in a cluster, we expect about 
4 times more arc(et)s systems in haloes corresponding to group masses, 
and 20 times more gravitational rings associated with lens galaxy
haloes (assuming equivalent detection limit and angular magnification 
factor in all cases) . We will see below that the 
preliminary number of group-lenses, based on the first (beta) version of
the arc detection software, is found to be $\sim0.5$\,deg$^{-2}$ in the CFHTLS 
data. Extrapolating Oguri's distribution to the total area of the CFHTLS 
Wide component (170 deg$^2$) yields $\sim75$ group arcs, 
$\sim400$ galaxy rings, and $\sim20$ cluster arcs. 
The number of lenses will be very small per Megacam field 
whatever the deflector mass regime (cluster, group or galaxy). 
Their detection is therefore very challenging as the lensed features are
hidden within a huge number of faint background galaxies.  The only way
to find them efficiently is to use well-defined automated procedures 
\citep[e.g. see ][]{lenzen04}. In practice such an approach has already 
been proposed for some large numerical simulations of lensing 
configurations, mostly in simulated samples of clusters of galaxies 
\citep{horesh05}.


\section{Automated software for Strong Lensing candidate 
detection\label{software}}
One of the goals of the SL2S is to provide a complete and
homogeneous sample of strong lenses with known detection efficiency with
regard to a series of observational parameters. This goal requires us not
only to develop a comprehensive series of simulations taking into account
the most common biases such as seeing, PSF variation, crowding, limiting
magnitude, surface brightness, and arc length/radius/thickness distribution,
but also requires us to automate the detection procedures. We are
thus currently developing three complementary algorithms covering
various regimes of strong lensing. The first one is a gravitational $ARC$
detector (mostly for groups and clusters).  The second is a detector of
compact $RING$ candidates (which will then require either spectroscopic
or higher resolution images for confirmations).  The third one is a
$MULTIPLET$ detector,  aimed at detecting peculiar and rare multiple
arclet systems that cannot be recognized by the $ARC$ detector. This
includes badly resolved multiple image systems or configurations which
may form in potential saddles.  The software selects multiple image configurations based only on generic properties
like color, distance, flux ratio and shear orientation, if any, and the
geometric distribution of multiple images.  However we have not yet been
able to optimize this procedure on ground-based images so the $MULTIPLET$
detector will not be addressed further.

\subsection{$ARC$ detector} 
Giant arcs are in principle the most straightforward images to identify
and detect through direct pattern recognition \citep{lenzen04,seidel06}.  
They are known to occur around massive 
clusters ($>10^{14} h^{-1} M_\odot$) and show radii 
up to $\sim$40\arcsec\ in very luminous X-ray 
clusters \citep[see for example the
spectacular case of A~1689][]{broadhurst05b}. When arcs appear around
groups of galaxies, their radii are five to ten times smaller and
the seeing makes them more difficult to identify from ground-based
imaging. A full description of the technique 
overcoming this difficulty is given in \citet{alard06}. 
Briefly, the arc-detector algorithm detects elongated
structures and analyzes the local properties of these structures.
Elongated objects are defined as very narrow objects along one
direction with a width nearly equal to the size of the seeing.
We model the elongated arc(let)s structures as small rectilinear 
objects having the width of the seeing along one
direction, and 3 times this width in the other direction.  This 
condition sets the detection threshold below an Einstein radius of 
3\arcsec\ for an arc angular aperture of about 60$^\circ$.  
Each object in the image is decomposed in a series of contiguous 
elements aligned along the tangential direction of the elongated structure. 
The full object is re-constructed by associating the areas covered 
by the different elements. Once a set of pixels is associated with the object, 
we compute its general properties, size, color, curvature, etc.  
To ease the pattern recognition the routine also requires both 
the $g'-i'$ color and surface brightness of the elongated 
objects to be constant.
Finally we produce a catalogue of candidates and a set of associated 
color images for visual inspection according to a selected set of 
parameters that fully describes the detection procedure.

The $ARC$ detector is very efficient at detecting extremely faint
arcs over a large range of splitting angles and will be 
used in subsequent CFHTLS releases (cf. Table~\ref{cfhtls_sl2s}).

\subsection{$RING$ detector}\label{ring_detector}

\begin{figure}
\includegraphics[width=9cm]{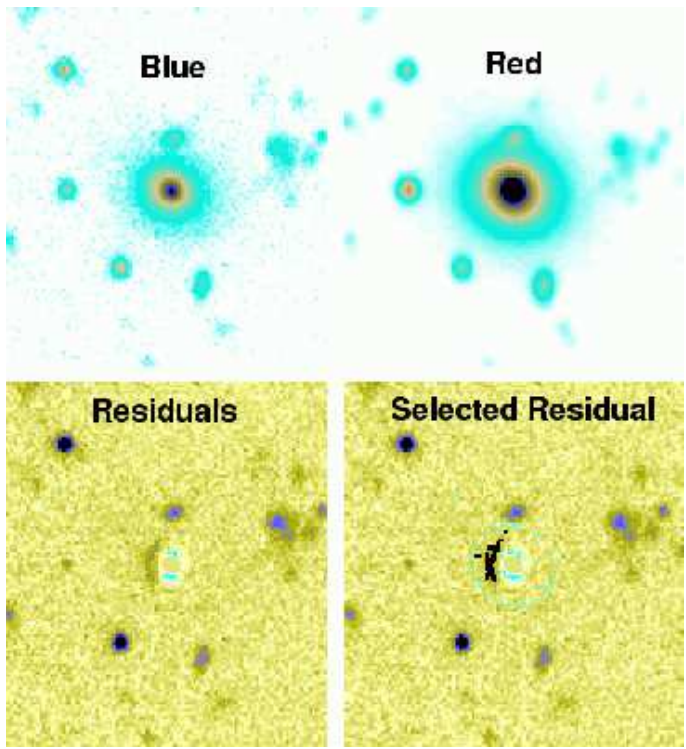}
\caption{Schematics of the ring detection procedure (here SL2SJ100013+022249). 
The candidates are selected from CFHTLS ellipticals (top). 
$g'-i'$ color images reveal strong residual signal (bottom). 
\label{ring_method}}
\end{figure}
\begin{figure*}
\includegraphics[width=\textwidth]{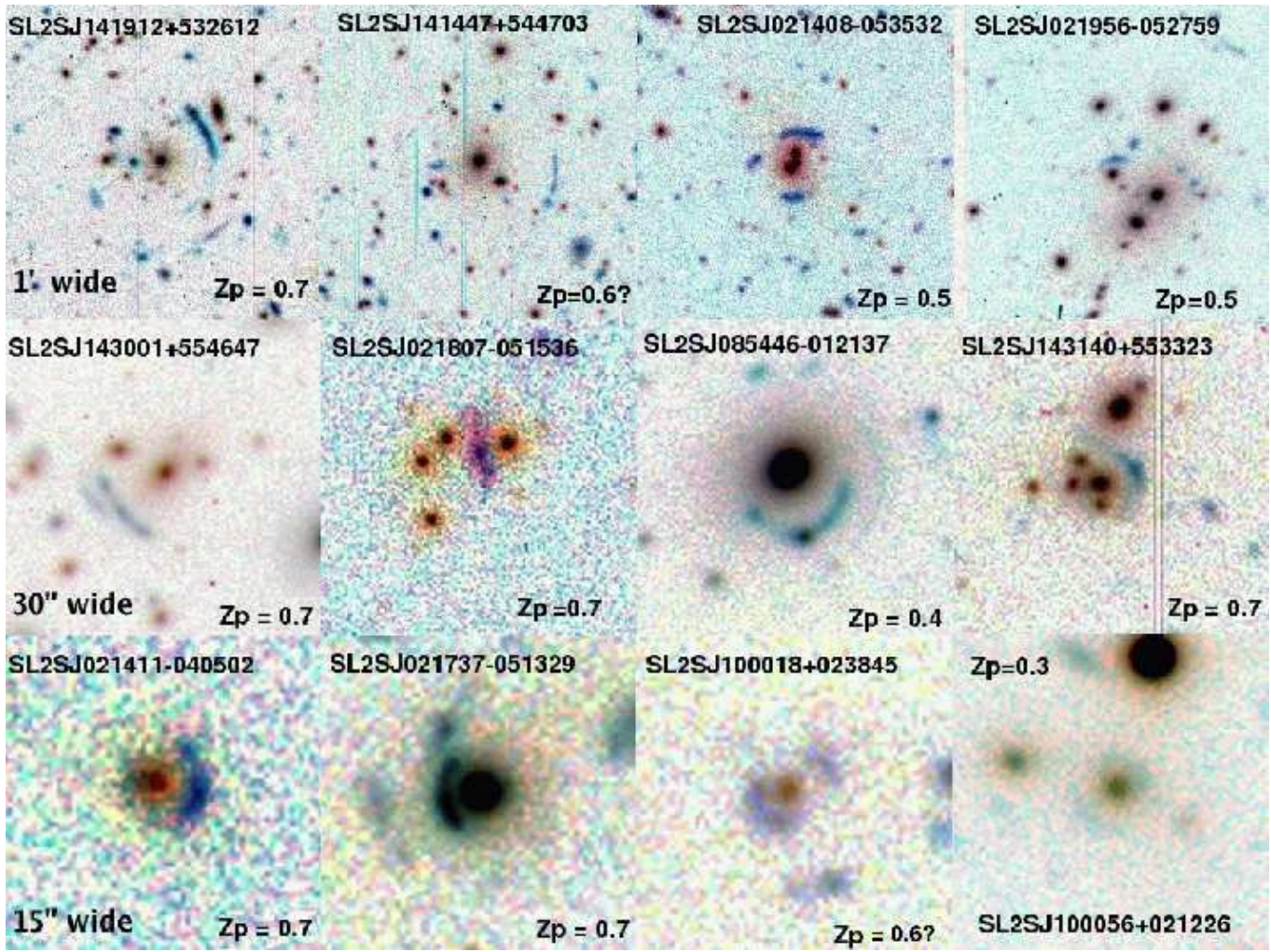}
\caption{\label{sl2s_mosaic} RGB ($i'g'u^*$) mosaic of SL2S candidates 
showing the three regimes probed by the sample. Top examples show
1\arcmin-wide 
vignettes of SL2S cluster lenses ($>10^{13} h^{-1} M_\odot$). Middle examples 
show 30\arcsec-vignettes of SL2S intermediate mass lenses (galaxy groups;
$\sim10^{13} h^{-1} M_\odot$). Bottom line shows 15\arcsec-wide vignettes of 
Einstein rings around single galaxy lenses ($<10^{13} h^{-1}
M_\odot$). The estimated photometric redshift for the deflector is marked.}
\end{figure*}
\begin{figure*}
\includegraphics[width=17cm]{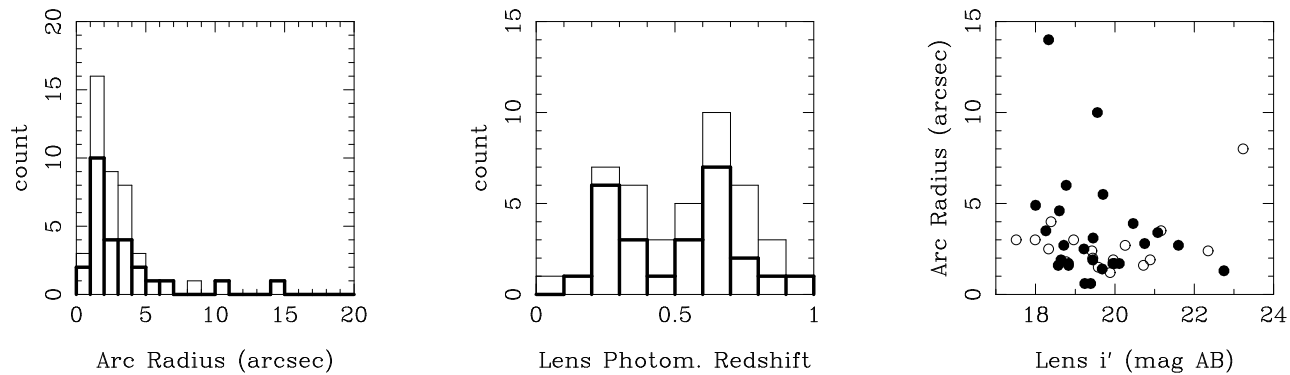}
\caption{T0002 release sample basic characteristics: Arc radius histogram 
(left), deflector/lens redshift histogram (middle). Arc radius 
versus central galaxy $i'$ magnitude of the deflector/lens (right).
Thick lines and filled circles correspond to quality 1 candidates. 
Thin lines and open circles correspond to quality 2 candidates (cf. text and Table~\ref{cfhtls_sl2s}). \label{sl2s_histo}}
\end{figure*}

The ring detector is aimed at detecting compact rings around centers
of isolated galaxies ($<10^{13} h^{-1} M_\odot$). Most of the ring
radii are in the range 0.5-2.0\arcsec\ and rings are usually hidden within
the deflector.  Seeing and intrinsic galaxy morphologies, like dust lanes 
and face-on spirals, make the ring detection challenging. 
We use an "object-oriented" routine, $RING$, that will be described 
in a future paper (Gavazzi et al., in preparation).
Presently, we focus on the 4 Deep fields that have been observed in
5 filters and for which photometric redshift catalogs are built.
These catalogs include a photometric redshift estimate, the best fit 
spectral type and information on the absolute magnitudes in different 
spectral bands. Selecting all objects catalogued as E/S0 galaxies, 
the routine filters out the large scale light distribution of the 
deflectors, using $g'-\alpha \times i'$ color vignettes. When the 
profile of the potential deflector does not depend much on color, 
or when the deflector profile is smooth on large scales, 
the deflector is subtracted cleanly and any residual comes from a 
superimposed smaller-scale anisotropy. The routine selects 
the lens candidates based on the computed residuals above sky noise 
in the range 0.8-2.5\arcsec (Fig.~\ref{ring_method}).  
We are currently optimizing the method with a sample of 10 
lens candidates common to the COSMOS field and the CFHTLS-D2 field. 
The first results are encouraging: most of the rings with radii larger 
than 0.8\arcsec\ seem to be recovered. However the method has two limitations.  
First, any ring candidate smaller than the seeing radius
is lost. Second, the procedure is a good filtering method, 
efficient in removing massive ellipticals that are not lenses, but
still produces false candidates among S0 galaxies. 
Therefore a final eyeball selection is currently still required
to select good candidates. 

The resulting candidates will need confirmation by spectroscopic 
identification of both the redshift of the lens and the lensed galaxy. 
In addition, high-resolution spectroscopy will help determine 
to the lens stellar velocity dispersion, 
and higher resolution images, using for example
the {\it Hubble Space Telescope}, will allow us to address 
the lens modelling with high accuracy. 
Thus, the method is mostly able to select, among a large number of 
massive ellipticals, a small sample of good strong lensing candidates. 

\subsection{Lens modelling}
A large number of algorithms of gravitational lens modelling are available
\citep[e.g.][]{kneib93,saha97,keeton01b,warren03,brewer06}.
It is clearly beyond the scope of this paper to review them. 
Our team has developped many parametric and non-parametric 
codes, using a variety of optimization algorithms to model the full 
spectrum of the mass regimes, therefore mostly adapted to the 
CFHTLS-SL2S sample.
 
\section{Preliminary sample from the T0002 release\label{sample}}
\subsection{Description of the sample}
The preliminary analysis of the CFHTLS T0002 release led to
the discovery of 43 candidate strong lenses listed in Table \ref{cfhtls_sl2s}.
Most of the CFHTLS-SL2S candidates that we present
here were extracted using the arc detector, which turned out to be
very efficient irrespective of the arc size or the lens environment.
The list also contains 10 rings first identified
in the COSMOS field and then recovered in the CFHTLS-D2
data.  Hence, this preliminary sample is considered 
a feasability study for the SL2S project, demonstrating that a 
wide mass range will be uncovered by the CFHTLS-SL2S. This sample
should not be used for quantitative statistical studies.

In the current selection process, the automated software
extracts a list of candidates and creates color images of each
of them for subsequent visual inspection. The SL2S database ({\tt
http://www.cfht.hawaii.edu/$\sim$cabanac/SL2S/}) is presently under
construction and includes for each system the magnitudes of the lens,
its photometric redshift estimates, using either the latest
version of the HyperZ code \citep{bolzonella00}, or LePhare, a code
based on a Bayesian approach \citep{ilbert06}.
A few geometrical properties for the lensed source (arc
radius, magnitudes when known) are also included. Table \ref{cfhtls_sl2s}
summarizes these measurements for our preliminary sample.

\subsection{Global properties of the sample}
Although this sample is not yet complete, we already see
three classes, split according to their arc radius, hence to their mass 
regime. A detailed mass classification will only be possible when spectroscopic 
redshifts of the lenses and the sources are known, and when high-resolution 
imaging {\bf yields} accurate lens modeling. 

\begin{figure}
\includegraphics[width=8.5cm]{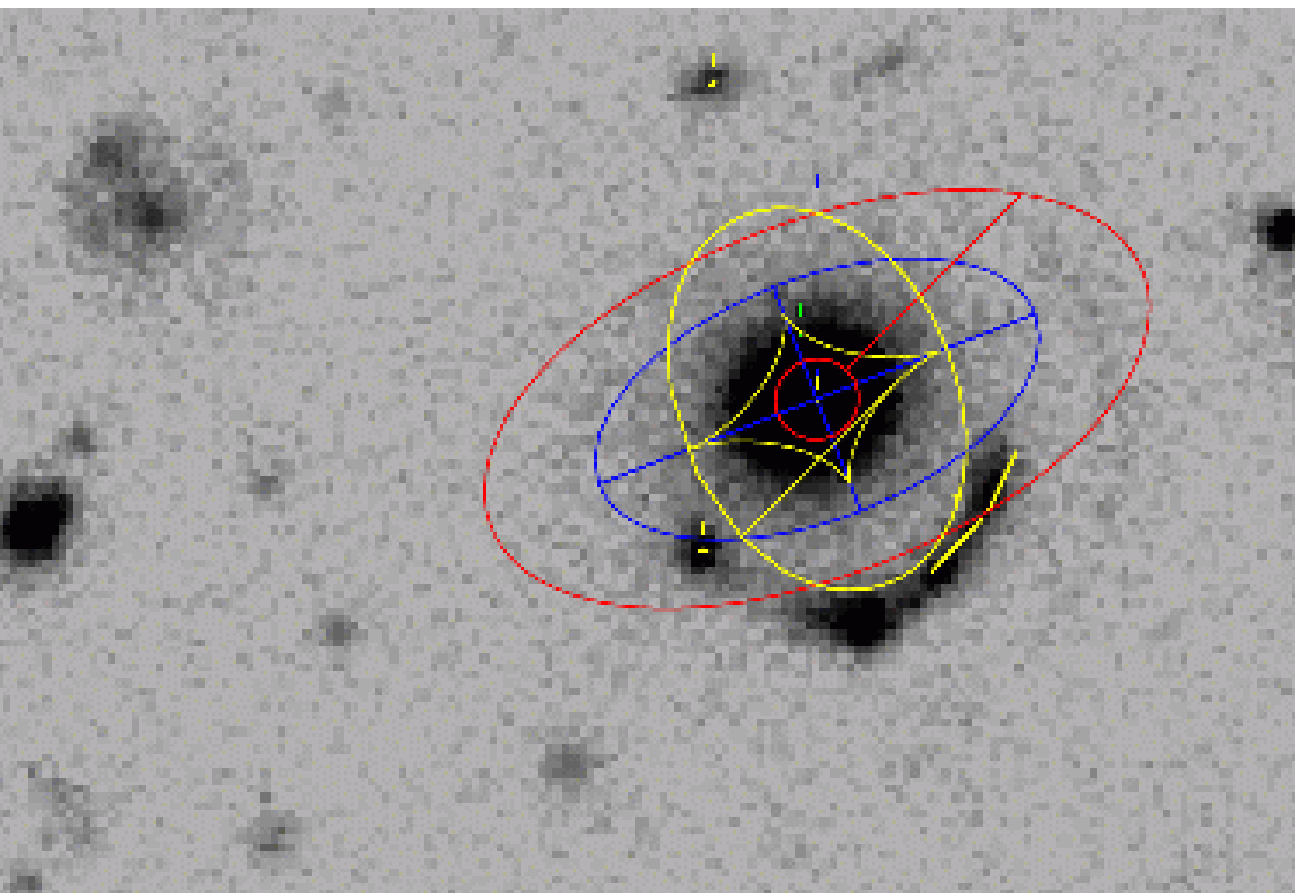}
\includegraphics[width=8.5cm]{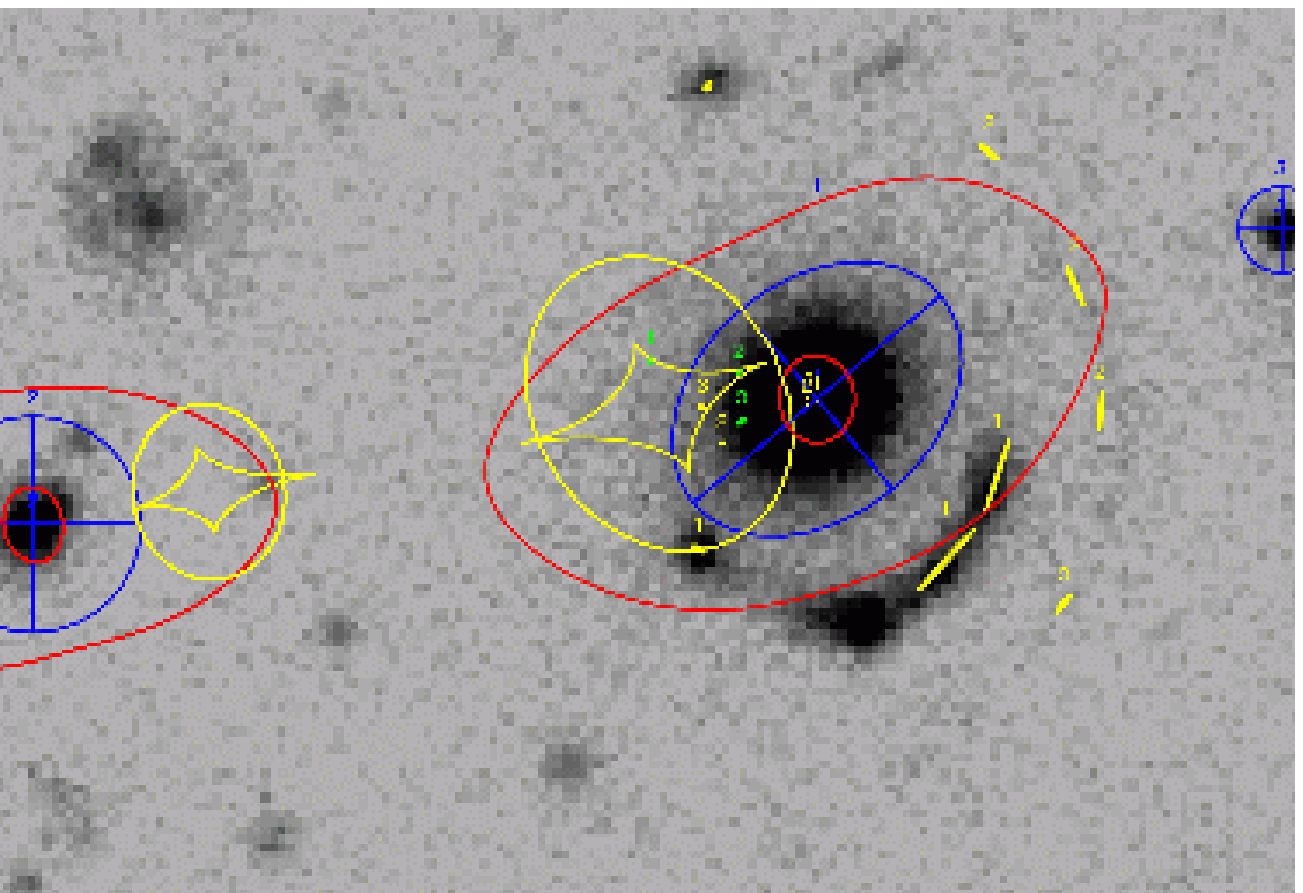}
\caption{(Top): Simple modelling of SL2SJ085446-012137 using LensTool
(Kneib 1993, http://www.oamp.fr/cosmology/lenstool/) where only the 
central galaxy is used as a deflector.
The source near the upper cusp of the radial critical line divides into
4 virtual images, three on the south half and one in the
north. (Bottom): Same lens with a more complex modelling including
surrounding galaxies as potential sources of shear. High-resolution
imaging and surface brightness information will allow one to remove
part of the degeneracy of lens models. \label{sl2s_model2}}
\end{figure}

\begin{itemize}
\item
The most conspicuous class includes giant arcs with radii $>7\arcsec $.
In the T0002 release sample, this class contains 4 candidates, detected
over an effective area of 28 sq. degrees. By simple extrapolation, one
can expect to find about $15-20$ of these giant arcs in the complete
CFHTLS 170 deg$^2$ (equivalent to about 100 deg$^2$ unmasked clear
sky).  The top line of the Figure~\ref{sl2s_mosaic} mosaic shows the 4 
giant arcs with arc radii $>7\arcsec $.  Usually such features 
appear in massive clusters often associated with strong X-ray emission.
\item
The second class of lenses is mostly made of intermediate mass
deflectors, showing arc radii in the range $3-7\arcsec $.  We found
13 such candidates (8 additional candidates with radii $\sim$
2.5\arcsec which might belong to 
small groups) in the CFHTLS T0002 release
($\sim 75$ are expected over the complete survey). The middle line
of Fig.~\ref{sl2s_mosaic} shows a selection of these intermediate mass
candidates. They represent the largest sample of intermediate mass lens
detected so far and the CFHTLS-SL2S seems to be particularly efficient
at detecting this class of lenses. The lens modeling of groups 
might be more complex than the modeling
of giant arcs in clusters of galaxies, often dominated by a bright
central galaxy, or of rings around an isolated elliptical galaxy. As an
illustration, Figure \ref{sl2s_model2} shows the possible influence of
the external shear due to group members on the main arc modeling for
the lens SL2SJ085446-012137. Only HST imaging will allow us to reduce
these degeneracies. 

\item
The third class is made up of compact ring candidates commonly associated
with isolated galaxies, with ring radii $<3\arcsec $.  This class is
expected to be the most populated one. If we follow Oguri's predictions
(2006) we expect the presence of about 100 rings in the T0002 Wide
survey release by a simple scaling between the intermediate mass lenses
and the galaxy mass lenses. If we exclude the 10 rings identified in the
COSMOS field (D2) using HST imaging and detected \emph{a posteriori} in the
CFHTLS data, there are only 12 extra candidates detected in
our sample. This shows that appropriate detection of ring candidates
is not yet fully operational and requires more developments and tests,
along the lines described in Section \ref{ring_detector}. The bottom line
of Fig.~\ref{sl2s_mosaic} shows a selection of the few compact arc or ring
candidates already identified.
\end{itemize}

The photometric redshift histograms of the deflectors of the total sample 
(thin line) and almost undisputable candidates (thick line) 
(Figure~\ref{sl2s_histo}) show a wide distribution up to $z \sim 1$. 
\citet{ilbert06} claim that LePhare Bayesian photometric redshifts
are accurate to $\sigma_{\Delta z} \sim 0.05$, with only 4\% of so-called catastrophic errors, i.e. when $\sigma_{\Delta z} > 0.15$, 
for objects brighter than $i_{AB}<24$, based on the 5 CFHTLS colors.
However, photometric redshifts of almost all the lenses detected in the 
Wide survey are determined
from 3 photometric colors only ($g', r', i'$) and should be taken with
caution, as the expected rate of catastrophic errors increases 
to $\sim$20\%.
The observed bimodal distribution with a peak about $z=0.3$ 
and another around $z=0.6$ could be a statistical fluctuation
or a real effect due to the fact that for the limiting magnitude 
of the CFHTLS Wide survey the distribution of sources
is expected to peak just above $z\sim0.6$.  A better determination of
the photometric redshifts, including near-IR data, or a spectroscopic
measurement through a dedicated spectroscopic follow-up are planned
to derive the observed redshift distribution of the sample.
With an average lens redshift $z\sim0.5$ the SL2S will go beyond the 
SLACS and will provide a galaxy and group sample able to probe 
the evolution of mass at higher redshift.
Fig.~\ref{sl2s_histo} also shows the expected loose
correlation between arc radius and lens apparent magnitude in $i'$ band, 
suggesting that the CFHTLS is not strongly biased 
in any mass regime, but seems, on the contrary, to be
sensitive to the complete parameter space.

\subsection{Notes on selected candidates}
\begin{figure}
\includegraphics[width=8.5cm]{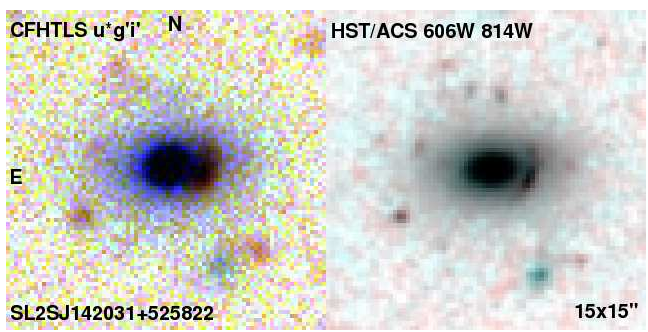}
\caption{(Left): CFHTLS RGB ($i'g'u^* $) image of the candidate 
SL2SJ142031+525822. (Right): HST RGB (814W-606W-600W) image of the 
same field clearly showing a bright arc akin to lensing distortion but 
a spectrum is needed to confirm that the arc is indeed at a higher redshift
and not a star-forming spiral arm or a merging galaxy.
\label{SL2SJ142031+525822}}
\end{figure}

\paragraph{SL2SJ021408-053532:} A bright system of fold arcs around a
compact group of three galaxies at $z_{phot} \sim 0.5$ (Fig.~\ref{sl2s_mosaic}), 
absolute $B$ magnitude $M_B=-22.4$, and a homogeneous population of 
galaxies with similar color within 2\arcmin.
\paragraph{SL2SJ021411-040502:} Typical example of a compact ring candidate
around a galaxy in a loose group or cluster at $z_{phot} \sim 0.7$.
These cases seem to be common in the CFHTLS and might provide interesting
constraints on the sub-structure of DM haloes.
\paragraph{SL2SJ100013+022249:} This is an example of a compact 
lensing candidate which does not appear very clearly in CFHTLS imaging, but
stands out in the HST cosmos field, and the CFHT normalized  $g'-i'$ image 
(Fig.~\ref{ring_method}).
\paragraph{SL2SJ141912+532612:} This massive cluster showing multiple arcs 
is not a new detection. \citet{gladders03} already discovered
the same lensing cluster called RCS1419.2+5326 within the Red Cluster Sequence 
survey. The photometric redshift of SL2SJ141912+532612 is $z_l=0.65$ 
based on 3 colors only ($g'r'i'$) but is remarkably similar to the spectroscopic 
redshift of RCS1419.2+5326 $z_{spectro}=0.64$. HST imaging in F814W is
available; HST proposal 10626, PI Loh. This system shows at least three arcs 
at 10, 14.5, and 17\arcsec (Fig.~\ref{sl2s_mosaic}).
\paragraph{SL2SJ142031+525822:} Example of a relatively compact arc/ring candidate 
at $z_{phot} \sim 0.3$ (and likely higher), the confirmation of which 
typically requires a spectroscopic follow-up in addition to
higher-resolution imaging. This candidate fortunately falls within the Groth 
Strip Survey and has already been imaged with HST/ACS in F606W and F814W 
\citep{simard02,vogt05}. Fig.~\ref{SL2SJ142031+525822} shows $1\times1\arcmin$ 
RGB images of the CFHTLS ($i'g'u^*$) and HST (814W-606W-600W). This candidate
is particularly interesting for the large ellipticity the deflector,
a rare occurence among isolated strong lenses.
\paragraph{SL2SJ142209+524652:} This candidate is a very extended arc 
surrounding a compact group of two galaxies at $z_{phot} \sim 0.2$ 
and offers a very unorthodox lensing configuration.
\paragraph{SL2SJ143140+553323} Example of lensing by two groups 
occupying the same line of sight at $z_{phot} \sim 0.5-0.6$, 
and having a discernable impact on the lensing distortion with a fold arc and 
a potential saddle pair.

\section{Discussion}
\subsection{Dark matter distribution in galaxy groups}
Only a handful of galaxy groups associated with strong lenses are
known, mostly as an environmental association with a galaxy lens
\citep{grant04,fassnacht05,auger06}. In such cases the group reveals 
its presence because it produces an external shear acting as a 
perturbation in the lens modeling. But so far almost no lenses with image
splitting between 3 and 7 arcsec have been observed corresponding to
halo masses of $M_{halo} \ga 10^{13} h^{-1} M_\odot$, except the
``historical'' double quasar Q 0957+561 \citep{walsh79}; this system
is centered in a giant elliptical galaxy embedded in a poor cluster
of galaxies at redshift 0.36 \citep{angonin94}, and recently
the 2 lensed quasars of the SDSS \citep{oguri05}.

There are several important reasons to study groups. 
60\% of the galaxies belong to groups which play a key role in the assembly
of structures.  Merging processes in groups can have a strong effect on
the star formation rate and probably on the growth of the super massive
black hole that lies in their dominant ellipticals. Much work
has been done in recent years using large surveys of groups in the
SDSS spectroscopic data base \citep{weinmann06}, within the DEEP survey
\citep{coil06} and/or with X-ray surveys \citep[ GEMS survey]{miles04}.
These first results did not succeed in giving a definitive picture 
of the average group mass profile, the relation between the
light and total mass distribution, or the relation between the different
galaxy type fractions (segregation effect as observed in clusters).

Part of the problem lies in the fact that sub-structures of 
simulated DM haloes can be very complex and the definition of a DM halo 
center is scale dependent. Strong lensing arcs in the center of group
haloes can provide a unique measure of the total projected density 
profile, where dynamical estimates only provide constraints on the baryonic
mass with the strong assumption that groups are virialized systems.

Recent weak shear analyses have been done by roughly
stacking two classes of groups detected in the CNOC2 spectroscopic
galaxy survey in order to derive structural and $M/L$ properties
\citep{moller02,parker05}. They found that very poor groups with a mean
velocity dispersion of 200\,km\,s$^{-1}$ have an average Einstein radius
of about 1\arcsec\ (very similar to single massive ellipticals) while the
most massive groups with a velocity dispersion close to 300\,km\,s$^{-1}$
have properties more similar to clusters. It also appears that a total
mass of 10$^{13} h^{-1} M_\odot$ might correspond to a transition
mass scale for the $M/L$ ratio and the related star formation rate.

X-ray observations have also revealed that there is a peculiar class 
of old groups dominated by a bright elliptical galaxy that probably formed at
an early time in the universe \citep{donghia05,ulmer05}. Likewise,
\citet{guimaraes05} used the QSO magnification bias in the 2dF Galaxy
Redshift Survey to measure a surprisingly high lensing signal, 
suggesting that some groups are more massive than expected. 


The SL2S group sample is well adapted to address these questions. 


\subsection{Searching a large sample of gravitational rings} 
Based on our simple extrapolations, CFHTLS Einstein rings will
clearly outnumber all the already existing galaxy lens samples. Once
fully optimised, the SL2S ring detection procedure will provide ca. 400
rings in the 170 deg$^2$ of the CFHTLS wide survey.  Even if we restrict
ourselves to a more conservative number of $\sim300$ robust SL2S systems,
confirmed with HST imaging, it will increase the present known sample by a
factor of 3 over a large redshift range $0.3-1$. Thus
a comprehensive statistical analysis of this class of lenses
will be possible, looking for variations as a function of cosmic 
time and galaxy luminosity, type and environment, 
and extending the results of the SLACS on the evolution of 
the E/S0 fundamental plane and the M/L ratio of normal galaxies,
following their pionneering method of combined spectroscopy-based
dynamical analysis and lens modelling.


\section{Conclusion}
We have presented the guidelines of the CFHTLS Strong
Lensing Legacy Survey, a survey aimed at detecting all the lensed features
in the CFHTLS data for the study of the dark matter halo distributions
from isolated galaxies to groups and large clusters up to a redshift of
1. A series of three automated softwares are being developed: 
an arc detector mainly focused on giant arcs and arclets,
a ring detector for the detection of compact Einstein rings and a
multiplet detector optimized in multiplet images showing generic lensing
properties irrespective of the deflector's light. The preliminary sample
is based on the CFHTLS T0002 release and led to the discovery of ca. 40
candidates spanning the complete mass range (The current release, T0003, 
increases the sample by 50\%. The best candidates will be followed up 
by HST in Cycle 15, snapshot program 10876). An unexpectedly large number
of intermediate mass deflectors akin to galaxy groups is present in this
sample. We therefore expect to address many questions related to galaxy
group physics through the project, like the mass profile transition
between cluster-like NFW and galaxy-like isothermal profiles,
dark matter substructures in intermediate mass haloes, star
formation processes and their feedback in the inter-galactic medium,
and stellar mass assembly with redshift.  Moreover, although the
detection procedure is not yet fully optimised, we expect to uncover
and study the most abundant population of rings, known to be
present in the CFHTLS images. This will allow us to extend the work of 
the SLACS team to higher redshifts for a better undertanding of the 
dark matter distribution and impact of baryon cooling in galaxy haloes, 
and the evolution of the Fundamental Plane and mass assembly of 
isolated galaxies.

By the end of the survey (2008), we expect to build a sample of about
500 lenses with well-controlled selection effects. This will allow us to
investigate the probability distribution of lens splitting angles across
the whole lens mass range from galaxies to clusters. Following the
pioneering work of \citet{kochanek01a}, the studies by \citet{ofek03}
and more recently \citet{oguri06} demonstrate how crucial it is to have
a large observational sample to be able to make meaningful comparisons
with simulations. The CFHTLS-SL2S is likely to become the most
complete survey of strong lenses available for many years.

\begin{acknowledgements}
The SL2S collaboration would like to thank P. Petitjean for sharing his
MegaCam dataset prior to publication, and C\'ecile Faure and the
COSMOS team for sharing their Einstein ring candidates in CFHTLS D2.
This work also uses Groth Strip Survey data obtained or processed with support 
of the National Science Foundation grants AST 95-29028 and AST 00-71198.
\end{acknowledgements}

\bibliography{sl2s}

\begin{sidewaystable*}
\begin{minipage}[t][180mm]{\textwidth}
\caption{CFHTLS-SL2S Terapix T0002 release (July 2005)\label{cfhtls_sl2s}}
\begin{tabular}{lcccccccccl}
\hline\hline
Name & RA (J2000) & DEC (J2000) & \multicolumn{5}{c}{AB mag$^a$} &
Arc radius$^b$ & $z_{lens}$ $^c$ & Comments$^d$\\
& hr~min~sec & $^\circ$~'~" & $u^*$ &$g'$ &$r'$ &$i'$ &$z'$ & arcsec & &\\
\hline
SL2SJ021258-051809& 02~12~58.114& -05~18~09.17& $21.86 \pm
0.02$&$20.73 \pm 0.01$&$19.51 \pm 0.01$ & $18.77 \pm 0.01$&$18.42 \pm
0.01$&$ 1.8 \pm0.2$ & 0.650 & W1 (2) arc\\
SL2SJ021301-043605& 02~13~01.862& -04~36~05.46& $23.93 \pm
0.07$&$21.47 \pm 0.01$&$20.29 \pm 0.01$ & $18.96 \pm 0.01$&$18.41 \pm
0.01$&$ 3 \pm0.5 $& 0.650 & W1 (2) arc\\
SL2SJ021308-053726& 02~13~08.693& -05~37~26.11& $21.96 \pm
0.03$&$20.12 \pm 0.01$&$18.87 \pm 0.01$ & $18.33 \pm 0.01$&$17.97 \pm
0.01$&$2.5 \pm0.3$ & 0.300 & W1 (2) vis\\
SL2SJ021311-041015& 02~13~11.261& -04~10~15.98& $24.98 \pm
0.31$&$23.82 \pm 0.09$&$23.10 \pm 0.07$&$22.35 \pm 0.04$&$22.14 \pm
0.09$&$ 2.4 \pm0.1$ & 0.725 & W1 (2) arc\\
SL2SJ021408-053532& 02~14~08.038& -05~35~32.30& $22.67 \pm
0.04$&$20.97 \pm 0.01$&$19.58 \pm 0.01$&$18.83 \pm 0.01$&$18.43 \pm
0.01$&$ 6 \pm1$ & 0.495 & W1 (1) arc\\
SL2SJ021411-040502& 02~14~11.212& -04~05~02.90& $23.02 \pm
0.06$&$22.20 \pm 0.02$&$21.04 \pm 0.01$&$19.95 \pm 0.01$&$19.54 \pm
0.01$&$ 1.7\pm 0.1$ & 0.735 & W1 (1) arc\\
SL2SJ021416-050315& 02~14~16.579& -05~03~15.51& $23.23 \pm
0.06$&$22.25 \pm 0.02$&$21.26 \pm 0.01$&$20.75 \pm 0.01$&$20.29 \pm
0.01$&$ 2.8 \pm0.3$ & 0.300 & W1 (1) arc\\
SL2SJ021613-061858& 02~16~13.991& -06~18~58.91& $22.61 \pm
0.06$&$21.92 \pm 0.02$&$20.97 \pm 0.02$ & $19.88 \pm 0.01$&$19.54 \pm
0.01$&$1.2 \pm0.1$ & 0.750 & W1 (2) arc\\
SL2SJ021737-051329& 02~17~37.232& -05~13~29.10& $22.23 \pm
0.03$&$21.37 \pm 0.01$&$20.62 \pm 0.01$ & $19.68 \pm 0.01$&$19.17 \pm
0.01 $&$1.4 \pm0.1$ & 0.67 & W1 (1) arc\\
SL2SJ021807-051536& 02~18~07.440& -05~15~36.16& $23.48 \pm
0.08$&$22.82 \pm 0.03$&$22.18 \pm 0.03$ & $21.60 \pm 0.02$&$21.10 \pm
0.04 $&$2.7 \pm0.5$ & 0.70 & W1 (1) arc\\
SL2SJ021932-053135& 02~19~32.029& -05~31~35.46& $26.08 \pm
0.79$&$23.97 \pm 0.06$&$22.24 \pm 0.03$ & $21.16 \pm 0.01$&$20.81 \pm
0.02 $&$3.5 \pm0.3$ & 0.755 & W1 (2) arc\\
SL2SJ021956-052759$^e$& 02~19~56.409& -05~27~59.08&$ 21.58 \pm
0.03$&$20.48 \pm 0.01$&$19.30 \pm 0.01$&$18.71 \pm 0.01$&$18.45 \pm
0.01 $&$2.7 \pm0.1$ & 0.5 & W1 (1) arc\\
SL2SJ022315-062904& 02~23~15.303& -06~29~04.92&$ 22.59 \pm
0.07$&$22.32 \pm 0.01$&$20.11 \pm 0.01$ & $19.22 \pm 0.01$&$18.91 \pm
0.01$&$ 2.5 \pm0.3$ & 0.95 & W1 (1) arc\\
SL2SJ022345-042402& 02~23~45.401& -04~24~02.16& $21.14 \pm
0.01$&$19.70 \pm 0.01$&$18.57 \pm 0.01$&$17.99 \pm 0.01$&$17.71 \pm
0.01$&$ 3 \pm1$ & 0.495 & W1 (2) arc\\
SL2SJ022532-045100 &02~25~31.990 &-04~51~00.22 &$21.60 \pm 0.01
$&$20.94 \pm 0.01$&$20.61 \pm 0.01$ & $19.44 \pm 0.01$ & $ 19.11 \pm
0.01$&$ 2.0 \pm 0.2 $& 0.495 & D1 (2) vis/ring\\
SL2SJ022914-065940& 02~29~14.265& -06~59~40.36&& $22.20 \pm
0.03$&$21.07 \pm 0.02$&$19.96 \pm 0.01$& &$ 1.9 \pm0.1$ & 0.900 & W1
(2) arc\\
SL2SJ023011-055023& 02~30~11.647& -05~50~23.64&& $19.65 \pm
0.01$&$18.17 \pm 0.01$&$17.51 \pm 0.01$& & $3.0 \pm0.5$ & 0.575 & W1
(2) arc\\
SL2SJ085446-012137$^f$&08~54~46.000 & -01~21~37.00& & $20.00 \pm
0.01$&$18.44 \pm 0.01$&$18.00 \pm 0.01$&&$4.9 \pm 0.2$& 0.35& W2 (1)\\
SL2SJ090407-005952& 09~04~07.915& -00~59~52.75&&$21.28 \pm
0.02$&$20.44 \pm 0.01$&$19.44 \pm 0.01$ && $1.9 \pm 0.1$& 0.750& W2
(1) arc\\
SL2SJ100009+022455 & 10~00~09.700&+02~24~55.00& $23.89 \pm 0.05$ &
$21.63 \pm 0.01$ & $20.22 \pm 0.01$ & $19.42 \pm 0.01$ &
$19.23 \pm 0.01$ &$2.4 \pm 0.2$& 0.505 &
D2 (2) arc\\
SL2SJ100012+022015 & 10~00~12.600&+02~20~15.00&$23.22 \pm
0.05$&$21.37 \pm 0.01$&$19.89 \pm0.01$ & $19.24 \pm 0.01$&$18.94 \pm
0.01$&$0.6\pm0.1$&0.284& D2 (1) vis\\
SL2SJ100013+022249&  10~00~13.900&+02~22~49.00& $22.58 \pm
0.05$&$20.55 \pm 0.01$&$19.19 \pm 0.01$& $18.57 \pm 0.01$&$18.29 \pm
0.01$&$1.6\pm 0.1$& 0.256& D2 (1) vis\\
SL2SJ100018+023845$^g$&  10~00~18.400&+02~38~45.00&$>26.1$&$25.68 \pm
0.28$&$23.59 \pm 0.05$&$22.75 \pm 0.03$ & $22.43 \pm 0.07$ &$ 1.3\pm
0.1$& 0.6?& D2 (1) vis\\
SL2SJ100056+021226&  10~00~56.700&+02~12~26.00&$22.69 \pm 0.04$&$20.75 \pm 0.01$&$19.30 \pm 0.01$&$	
18.64 \pm 0.01$&$18.32 \pm 0.01$& $1.9 \pm0.1$& 0.256& D2 (1) vis\\
SL2SJ100148+022325&10~01~48.100&+02~23~25.00& $22.98 \pm 0.05$&$20.97
\pm 0.01$&$19.47 \pm 0.01$&$18.83 \pm 0.01$&$18.52 \pm 0.01$&$1.6\pm
0.2$& 0.284& D2 (1) vis\\
SL2SJ100208+021422& 10~02~08.500& +02~14~22.00&$23.90 \pm
0.08$&$22.03 \pm 0.01$&$20.63 \pm 0.01$&$19.99\pm 0.01$&$19.65 \pm
0.01$ &$ 1.7\pm 0.2 $&0.256 &D2 (1) vis\\
SL2SJ100211+021139$^g$& 10~02~11.200& +02~11~39.00&$25.56 \pm
0.33$&$24.12 \pm 0.07$&$22.37 \pm 0.02$ & $21.08\pm 0.01$&$20.29 \pm
0.01$ &$ 3.4 \pm 0.2$& 0.54 &D2 (1) vis\\
SL2SJ100216+022955& 10~02~16.800& +02~29~55.00& $27.05 \pm
1.27$&$22.88 \pm 0.03$&$21.28 \pm 0.01$& $20.11 \pm 0.01$&$19.68 \pm
0.01$&$ 1.7 \pm0.2 $&0.52 &D2 (1) vis\\
SL2SJ100221+023440& 10~02~21.100& +02~34~40.00& $23.57 \pm
0.06$&$21.57 \pm 0.01$&$20.06 \pm 0.01$& $19.39 \pm 0.01$&$19.05 \pm
0.01$&$0.6\pm0.2$ & 0.256 &D2 (1) vis\\
SL2SJ141447+544703$^h$ & 14~14~45.498& +54~47~00.03 & &$21.35 \pm 0.03$&
$19.59 \pm 0.01$ & $18.33 \pm 0.01$&& $ 14 \pm1$& 0.75&
W3 (1) vis/arc\\
SL2SJ141558+523955& 14~15~58.182& +52~39~55.92&& $21.45 \pm 0.01$& 
$19.77 \pm 0.01$ & $18.60 \pm 0.01$&& $ 4.6 \pm0.2$& 0.75 & W3 (1) vis/arc\\
SL2SJ141807+524924& 14~18~07.966&+52~49~24.28&$22.42 \pm 0.04$&$20.43
\pm 0.01$&$19.00 \pm 0.01$&$18.39 \pm 0.01$&$18.08 \pm 0.01$&$4\pm
0.5$& 0.256& D3 (2) vis\\
\hline
\end{tabular}

$^a$ AB mag as given in the official release, Sextractor Best Magnitudes, (AB to Vega $u^*-0.35$, $g'+0.09$, $r'-0.17$, $i'-0.40$, $z'-0.55$).\\
$^b$ Arc radii are computed from LensTool program (Kneib 2001).\\
$^c$ $z_{lens}$: Best photometric redshift given by HyperZ, taking into account all available filters and $E(B-V)$ (Pello 2006, priv. comm.).\\
$^d$ Comments: CFHTLS component, Numbers refer to: (1) Good quality candidate based on visual inspection (2) Potential candidate based on visual inspection, Detection method: arc = arc detector, vis = visual, ring = ring detector.\\
$^e$ The bright arc of this candidate is around a 
secondary peak (Fig.~\ref{sl2s_mosaic}). If confirmed, it is a good example of an enhanced convergence in a dense field with presence of sub-haloes.\\
$^f$ Massive cluster lens.\\
$^g$ Compact and red candidate ( $u$ dropouts) in the field of COSMOS. 
Its photometric redshift is not well-constrained because the source contaminates the deflector. For SL2S100018, the source has a best $z_{phot}\sim4.2$. \\
$^h$ Massive system with a bright radial arc.\\
\end{minipage}
\end{sidewaystable*}
\begin{sidewaystable*}
\begin{minipage}[t][180mm]{\textwidth}
{\bf Table~\ref{cfhtls_sl2s}.}~Continued.\\[12pt]
\begin{tabular}{lcccccccccl}
\hline\hline
Name & RA (J2000) & DEC (J2000) & \multicolumn{5}{c}{AB mag$^a$} &
Arc radius$^b$ & $z_{lens}$ $^c$ & Comments$^d$\\
& hr~min~sec & $^\circ$~'~" & $u^*$ &$g'$ &$r'$ &$i'$ &$z'$ & arcsec & &\\
\hline
SL2SJ141912+532612$^i$& 14~19~12.000& +53~26~12.00&& $22.27 \pm
0.02$&$20.94 \pm 0.01$&$19.56 \pm 0.01$&& $10-17\pm 0.5$ &0.65 &W3 (1) vis/arc\\
SL2SJ142028+521303& 14~20~28.799&+52~13~03.54& $22.51 \pm
0.03$&$21.61 \pm0.01$&$20.65 \pm 0.01$ & $20.26 \pm0.01$ & $20.02 \pm
0.01$ & $2.7 \pm0.3$ & 0.344& W3 (2) arc\\
SL2SJ142031+525822$^j$& 14~20~31.811&+52~58~22.16&$ 22.53 \pm
0.04$&$20.95 \pm 0.01$&$19.48 \pm 0.01$&$18.83 \pm 0.01$&$18.48\pm
0.01$&$1.7\pm 0.2$& 0.300& D3 (1) vis\\
SL2SJ142032+530107& 14~20~32.147&+53~01~07.13& $22.98 \pm
0.04$&$22.20 \pm 0.01$&$21.58\pm 0.01$&$20.72 \pm 0.01$&$20.44 \pm
0.01$&$1.6\pm 0.2$& 0.74& W3 (2) vis\\
SL2SJ142057+530843$^j$&14~20~57.684&+53~08~43.81&$>25.8$&$>26.3$&$25.92
\pm 0.53$&$23.23\pm 0.06$&$23.42 \pm 0.26$&$8.0\pm 1.0$ &0.622&
D3 (2) vis\\
SL2SJ142207+531013& 14~22~07.611&+53~10~13.88& $21.46 \pm
0.01$&$20.64 \pm 0.01$&$19.87 \pm 0.01$&$19.58 \pm 0.01$&$19.40 \pm
0.01$&$1.5\pm 0.2$& 0.310& D3 (2) ring\\
SL2SJ142209+524652&14~22~09.268& +52~46~52.42&$21.15 \pm 0.01$&$19.87
\pm 0.01$&$18.75 \pm 0.01$&$18.26 \pm 0.01$&$17.99 \pm 0.01$&$3.5\pm
0.5$& 0.189& D3 (1) vis\\
SL2SJ142258+512440& 14~22~58.343& +51~24~40.79&& $22.83 \pm
0.02$&$21.77 \pm 0.03$&$20.89 \pm 0.01$&& $1.9 \pm0.1$& 0.87 &W3 (2) arc\\
SL2SJ143001+554647$^k$& 14~30~01.000& +55~46~47.00&& $21.93 \pm
0.02$&$20.59 \pm 0.01$&$19.70 \pm 0.01$&& $5.5\pm 1 $&0.695& W3 (1) vis/arc\\
SL2SJ143140+553323&14~31~40.000 & +55~33~23.00&&$21.81 \pm
0.02$&$20.09 \pm 0.01$&$19.45 \pm 0.01$&&$3.1 \pm0.7$ &0.655& W3 (1) vis\\
SL2SJ143141+513143& 14~31~41.829& +51~31~43.71&& $22.86 \pm
0.02$&$21.38 \pm 0.02$&$20.46 \pm 0.01$&& $3.9\pm 0.2$ &0.85& W3 (1) arc\\
\hline
\end{tabular}

$^a$ AB mag as given in the official release, Sextractor Best Magnitudes, (AB to Vega $u^*-0.35$, $g'+0.09$, $r'-0.17$, $i'-0.40$, $z'-0.55$).\\
$^b$ Arc radii are computed from LensTool program (Kneib 2001).\\
$^c$ $z_{lens}$: Best photometric redshift given by HyperZ, taking into account all available filters and $E(B-V)$ (Pello 2006, priv. comm.).\\
$^d$ Comments: CFHTLS component, Numbers refer to: (1) Good quality candidate based on visual inspection (2) Potential candidate based on visual inspection, Detection method: arc = arc detector, vis = visual, ring = ring detector.\\
$^i$ Other name RCS1419.2+5326 \citep[$z_{spectro}=0.64$; ][]{gladders03}, HST imaging available in F814W; 
HST proposal 10626, PI Loh. This system shows at least three arcs 
at 10, 14.5, and 17\arcsec.\\
$^j$ HST imaging available in F606W and F814W from the Groth Strip Survey 
\citep{simard02,vogt05}, probably a tidal tail.\\
$^k$ Massive cluster lens with fold arc.\\

\end{minipage}
\end{sidewaystable*}
\end{document}